\documentclass[10pt,prl,aps,twocolumn,showpacs,superscriptaddress]{revtex4}
\usepackage{graphicx}
\usepackage{amssymb}
\usepackage{bm}

\newcommand{\bc}{\begin{center}}
\newcommand{\ec}{\end{center}}
\def\ba#1{\begin{array}{#1}\displaystyle}
\newcommand{\ea}{\end{array}}

\newcommand{\beq}{\begin{equation}}
\newcommand{\eeq}{\end{equation}}
\newcommand{\beqa}{\begin{eqnarray}}
\newcommand{\eeqa}{\end{eqnarray}}

\newcommand{\n}{\nonumber\\}
\newcommand{\bi}{\begin{itemize}}
\newcommand{\ei}{\end{itemize}}
\def\mato#1{\left(\ba{#1}} 
\def\matf{\ea\right)}

\def\lt#1{\left#1}
\def\rt#1{\right#1}

\def\h#1{\hat{#1}}

\def\frc#1#2{\frac{#1}{#2}}

\newcommand{\bra}{\langle}
\newcommand{\ket}{\rangle}
\newcommand{\Z}{{\mathbb{Z}}}

\newcommand{\R}{{\mathbb{R}}}

\newcommand{\Tr}{{\rm Tr}}

\newcommand{\Or}{{\cal O}}

\newcommand{\ep}{\epsilon}

\begin{document}

\title{Entanglement entropy of highly degenerate states and fractal dimensions}

\author{Olalla A. Castro-Alvaredo}
\affiliation
{Centre for Mathematical Science, City University London, Northampton Square EC1V 0HB, U.K.}
\author{Benjamin Doyon}
\affiliation
{Department of Mathematics, King's College London, Strand WC2R 2LS, U.K.}

\date{\today}


\begin{abstract}
We consider the bi-partite entanglement entropy of ground states of extended quantum systems with a large degeneracy. Often, as when there is a spontaneously broken global Lie group symmetry, basis elements of the lowest-energy space form a natural geometrical structure. For instance, the spins of a spin-1/2 representation, pointing in various directions, form a sphere. We show that for subsystems with a large number $m$ of local degrees of freedom, the entanglement entropy diverges as $\frac{d}{2}\log m$, where $d$ is the fractal dimension of the subset of basis elements with non-zero coefficients. We interpret this result by seeing $d$ as the (not necessarily integer) number of zero-energy Goldstone bosons describing the ground state. We suggest that this result holds quite generally for largely degenerate ground states, with potential applications to spin glasses and quenched disorder.
\end{abstract}

\maketitle

The entanglement entropy is a measure of entanglement  between two
complementary sets of observables in a quantum system
\cite{bennet}. It is defined as the von Neumann entropy
of the
reduced density matrix of the state $|\Psi\ket$ with
respect to a tensor factor of the Hilbert space ${\cal H}$: \beq
\label{def}
    S = -\Tr_{\cal A}(\rho_A\,\log\rho_A)\quad \text{with} \quad \rho_A = \Tr_{\cal B}|\Psi\ket\bra\Psi|,
\eeq and ${\cal H} = {\cal A} \otimes {\cal B}$. A related measure is obtained from the R\'enyi  entropy, $S_n = \frc1{1-n}\log \Tr_{\cal A}(\rho_A^n)$;
clearly, $S = S_1=\lim_{n\to 1^+}S_n$. The entanglement and R\'enyi entropies have important applications to e.g. quantum computation and numerical simulations of quantum systems.

In extended quantum systems near to critical points,  the
entanglement entropy has turned out to reveal fundamental
properties of ground states (for reviews, see e.g. \cite{special}). An important result is the so-called
area law. Consider a quantum system of dimensionality $D\ge 2$
with correlation length $\xi$, and a subsystem ${\cal A}$ composed
of the local degrees of freedom on a $D$-dimensional region $A$ of
linear extension $\ell$ (generically, the region $A$ is composed various components of different connectivities, and $\ell$ is the overall scale of $A$). It turns out that the entanglement
entropy between the subsystem and the rest diverges as $\xi$ and
$\ell$ increase, the ratio $r=\ell/\xi$ being fixed, with a power
law $\ell^{D-1}$, with possible logarithmic corrections for gapless systems \cite{Bombelli:1986rw,Srednicki:1993im,Wolf08,Wolf06,Gioev06,CalMinVic11}. But this area law is special in the cases
where $D=1$. There, the divergence is always logarithmic:
$\frc{qc}6\log(\ell)$ where $q$ is the number of points separating $A$ from the rest
and $c$ is the central charge of the critical theory
\cite{Holzhey,CalabreseCardy}. Interestingly, the
number $c$ comes out, which essentially measures the number of
degrees of freedom that are carried over from the microscopic
theory to the macroscopic universal theory. Further, for $D=1$ again, subtracting this
divergence, the rest is a finite quantity which depends on $r$, which saturates to a finite value
at $r=\infty$, and which tends to this value in an exponential way
that is solely determined by the spectrum of masses of the
corresponding perturbation of the critical point \cite{entropy,entropy2}. The spectrum of asymptotic
particles characterizes the low-energy degrees of freedom of the
universal theory. Moreover, in systems with a boundary, the
boundary degeneracy appears also by a natural subtraction \cite{CalabreseCardy,entropy3}. This
degeneracy characterizes the number of degrees of
freedom carried by the boundary. These results point to the
observation that if the entanglement entropy diverges
logarithmically at large subsystem size $\ell$, then the way it
diverges is controlled by some basic counting of universal degrees of
freedom.

All these results were established for non-degenerate ground states, or ground states with small, finite degeneracies. A question arises as to the entanglement entropy for highly-degenerate ground states. Let us start by discussing an example where a symmetry group is spontaneously broken: the Heisenberg ferromagnet. This is an $N$-site lattice ${\tt L}$ with spin-$1/2$ local degrees of freedom, with Hamiltonian: $H = J\sum_{(i,j)\;\in\; {\rm edges\ of\ }{\tt L}} \vec\sigma_i\cdot\vec\sigma_j,\quad J<0\label{xxx}$ ($\vec\sigma_i$ is a vector of Pauli matrices acting on  site
$i$). This model has an $SU(2)$ global symmetry, and the states
$|\psi_{\vec{v}}\ket^{(N)}=\otimes_{i\in{\tt L}} |\psi_{\vec{v}}\ket_i$,
where all spins point in the same direction $\vec{v}$ (i.e.~$\vec\sigma_i\cdot\vec{v}|\psi_{\vec{v}}\ket_i =
|\psi_{\vec{v}}\ket_i$, $|\vec{v}|=1$), span the
lowest-energy subspace. In the usual description, we make a choice of an arbitrary direction
$\vec{v}_0$. Such a state is not
invariant under $SU(2)$ transformations, hence the symmetry is
dynamically broken. The Hilbert space ${\cal H}_{\vec{v}_0}$ in the 
thermodynamic limit $N\to\infty$ is then, with the ground state
$|\psi_{\vec{v}_0}\ket^{(N)}$, the set of all
finite-energy, local excitations above it. By
locality of the Hamiltonian, it excludes all ground
states and excited states associated to other directions, ${\cal
H}_{\vec{v}}$ for $\vec{v}\neq \vec{v}_0$ (these cannot be reached by a finite number of local changes of the infinite system).

But linear combinations of $|\psi_{\vec{v}}\ket^{(N)}$s also give
lowest-energy states, and for them we will take a different description of quantum states that is more appropriate.
The lowest-energy subspace is the
$N+1$-dimensional subspace forming a spin-$N/2$ representation. For every $N$, there exists a set of $N+1$ points $\vec{v}_k$ such that the set of vectors $|\psi_{\vec{v}_k}\ket^{(N)}$ forms a basis for this subspace. Further, in the limit $N\to\infty$, every point on the unit sphere is arbitrarily close to such a basis point. A good description of the resulting space of infinite-volume lowest-energy quantum states is then obtained by using the geometry induced by averages of local operators (see \cite{permutation}, Section 4).
In this geometry, the
distance between states in directions $\vec{v}$ and $\vec{v}'$ is
smoothly related to the distances between the vectors $\vec{v}$
and $\vec{v}'$ on the unit sphere. This geometry is convenient for
discussing the entanglement entropy, because, as is developed in
\cite{permutation} (based on earlier works \cite{entropy}),
the latter can be evaluated from the average of a local
permutation operator.

Linear combinations could involve infinitely  many
directions $\vec{v}$, with appropriate integration measures on the
unit sphere. This occurs, e.g., when a ground state of the infinite-length
one-dimensional Heisenberg ferromagnet is reached by an adiabatic lowering
of the anisotropy of the XXZ model: an integration
over a great circle on the unit sphere is obtained \cite{permutation}. Although each state $|\psi_{\vec{v}}\ket^{(\infty)}$ has zero entanglement
entropy (since it is factorisable), linear combinations do not,
and linear combinations involving infinitely many directions
$\vec{v}$ should have growing entropy as
$\ell\to\infty$. What is the $\ell\to\infty$ behaviour for such infinite linear combinations?

Let ${\cal A}$ be composed of $m$ degrees
of freedom and $N=\infty$.
Clearly, any minimal-energy state has a symmetry under exchange of
any two sites, hence the entanglement entropy depends on $m$ but
not on the particular sites chosen. We may take the $m$ sites to
form a continuum of dimension $D$, writing $m =
\ell^D$. First note that a large-$m$ divergence $\frc{1}2\log m$
of the entanglement entropy was found in \cite{permutation,popkov} for
the state formed by an integration over a great circle. Second,
recall that when there is spontaneous symmetry breaking, there are
massless excitations in the spectrum, the Goldstone bosons. In
general, our idea is that linear combinations composed of all points along an
arc on the unit sphere should be interpreted as representing the
presence of a zero-energy Goldstone boson corresponding to the
continuous motion along this arc. Moreover, every linearly
independent local direction on the unit sphere corresponds to a
linearly independent Goldstone boson, each of which can be seen as
a universal degree of freedom. Hence, the observations above
suggest a divergence of the
form $\frc{d}2 \log m$, where $d$ is the number of Goldstone
degrees of freedom present in the linear combinations. This number
is simply the dimension of the support of the linear combination
on the unit sphere. The ``number'' of Goldstone
degrees of freedom $d$ is not restricted to the integers: the
support of the linear combination may have a fractal
dimension.

Here we argue that the
large-$m$ (large-$\ell$) behaviour is
\beq\label{main}
    S_n = \frc{d}2 \log m + O(1) = \frc{dD}{2} \log \ell + O(1)
\eeq for all $n$, where $d$ is the (fractal) dimension of the support  of the
linear combination, $0\leq d\leq 2$ for the spin-$1/2$ Heisenberg ferromagnet. For instance, if the great circle in the above example is replaced by the Cantor set, we would find $d=\log(2)/\log(3)$. Note that the result is independent of $n$.

A simple application of this formula is to detect a possible blurring of  the dynamically chosen direction $\vec{v}_0$ obtained, for instance, as the system is cooled in a fixed magnetic field. The blurring could lead to a linear combination covering a possibly fractal small surface around $\vec{v}_0$. Standard local observables would not discern this, whereas formula (\ref{main}) shows that the entanglement entropy is extremely sensitive to it, giving $d>0$ instead of $d=0$.

Beyond the Heisenberg ferromagnet, our derivation below makes it clear
that (\ref{main}) should hold much more generally. There are
two conditions: 1) there exists a basis for the lowest-energy
subspace where the entanglement entropy of each basis element is
zero or small, and 2) the basis elements are given
the local-operator geometry \cite{permutation}. The dimension $d$ in (\ref{main}) is
that of the support of the linear combination in this geometry, which may be an integer or not, and which essentially counts the number of Goldsone bosons in the state. For instance,
for a quantum system in a ``Mexican hat'' potential, the
basis set is the geometrical circle at the bottom of the hat, and subsets of this will have $0\leq d\leq 1$; similar observations hold for any system with spontaneously broken continuous symmetry.

Note that some ``permutation-symmetric'' (PS) states of the above type in a Hilbert space with on-site spin $S$ were considered in \cite{popkov}, and (\ref{main}) with $d=2S$ was observed. An analysis shows that the states chosen generalize the $S=1/2$ great-circle state. Further, this is in agreement with our general arguments, which imply that all possibilities $0\leq d\leq 4S$ may occur for PS states.

Our formula is in sharp contrast with behaviors reviewed above (e.g. $\ell^{D-1}$  for $D>1$), related to the geometric structure of the region $A$ and arising thanks to locality of the system's interaction. To explain this, consider the case of a spontaneous symmetry breaking: local interactions only fix the lowest-energy subspace, not the ground state. By choosing a basis of lowest-energy states with minimal entanglement and with a local-operator geometry, we expect that we encode all local information in the basis states themselves, and only the symmetry is probed by the degeneracy. Hence, our result, which has to do with the degeneracy, cannot measure geometrical objects in the system's real space. Rather, a logarithm occurs, whose coefficient counts Goldstone degrees of freedom, associated with the symmetry. If condition 1) above does not hold, we expect two contributions to the asymptotic of the entanglement entropy: that of the large degeneracy (\ref{main}), and that coming from locality.

Examples of fractal sets of minima are found wherever random potentials occur (quenched disorder, see e.g. \cite{CarLeDou}), with possible connections to glasses. In the classical phenomenology \cite{Dotsenko}, at temperatures below the glass transition point, the free energy surface reveals finer structures in the form of new energy minima within previous valleys, displaying self-similarity and a fractal structure; the set of effective minima has a nontrivial fractal dimension. High classical degeneracy also naturally occurs in frustrated spin systems \cite{Diep}. These degeneracies may be lifted by quantum fluctuations (so-called ``order by disorder''), although the underlying classical degeneracy is known to have nontrivial quantum effects and to survive semi-classically \cite{Berg}. By our formula (\ref{main}), the entanglement entropy could provide a further indicator at the quantum level of this (semi-)classical degeneracy.

In the rest of this paper, we provide the main lines of the
derivation of (\ref{main}).  A more precise proof and statement
will be given in a separate work.

\section{The Heisenberg ferromagnet case}

The present derivation uses the replica trick, whereby $n$ is assumed to be an integer $>1$. The result, however, can be interpreted for $n\in(1,\infty)$. This provides the unique analytic continuation which does not diverge exponentially at large $n$; as is usual, this analytic continuation is assumed to provide $S_n$ for all real $n\geq1$.

Given a point $\vec{v}$ on the unit sphere $S^2$, let us denote by
$\psi_{\vec{v}}\in{\cal F}$ the quantum state corresponding to
the $N\to\infty$ limit of $|\psi_{\vec{v}}\ket^{(N)}$. As
developed in \cite{permutation}, a quantum state is
a linear functional on the space of finitely-supported operators,
which evaluates the average; e.g.
$\psi_{\vec{v}}(\Or) = \lim_{N\to\infty}{}^{(N)}\bra\psi_{\vec{v}}|\Or|\psi_{\vec{v}}\ket^{(N)}$.
We can write $\psi_{\vec{v}}$ as a product of single-site
quantum states, all acting in the same way: $
    \psi_{\vec{v}} = \bigotimes_{i\in\Z} \psi_{\vec{v};i}.$
At infinite volume, vectors pointing in different directions have
zero overlap:
$\lim_{N\to\infty} {}^{(N)}\bra
\psi_{\vec{v}}|\psi_{\vec{v'}}\ket^{(N)} = 0$ for $\vec{v}\neq
\vec{v'}$. This holds as well with insertions  of
finitely-supported operators, so the infinite-volume limit of
linear combinations $\sum_{\vec{v}} a_{\vec{v}}
|\psi_{\vec{v}}\ket^{(N)}$ gives the quantum state
\beq\label{pa}
    \psi_{\{a_{\vec{v}}\}} := \sum_{\vec{v}} |a_{\vec{v}}|^2 \psi_{\vec{v}},\quad
    \sum_{\vec{v}} |a_{\vec{v}}|^2 = 1.
\eeq

In order to evaluate the R\'enyi entanglement entropy associated
to the ground state $\psi_{\{a_{\vec{v}}\}}$ we recall the approach developed in
\cite{permutation}. There, the R\'enyi entropy of a region $A$ in
a quantum state $\psi$ was expressed as an average on the
$n^{\rm th}$ tensor power of $\psi$:
\begin{equation}\label{ren}
    S_n =\frc1{1-n} \log\left(\psi^{\otimes n}(
    \mathcal{T}_A)\right).
\end{equation}
The operator averaged is $\mathcal{T}_A=\prod_{i \in A}
\mathcal{T}_i$,  where $\mathcal{T}_i$ are \emph{local cyclic
replica permutation operators} which act on site $i$ of the
quantum spin chain by cyclicly permuting the spins of the $n$
replicas of the model at that particular site. One of the
results of \cite{permutation} was the closed formula
\begin{equation}\label{trace}
    {\mathcal{T}}_i=\text{Tr}_{\text{aux}}
        \prod_{\alpha=1}^n \sum_{\ep_1,\ep_2} E^{\ep_1\ep_2}_{\text{aux}}
        E_{\alpha,i}^{\ep_2\ep_1},
\end{equation}
where $ E_{V}^{\ep_2\ep_1}$ represent elementary $2\times
2$  matrices with a single non-vanishing entry at row $\ep_2$,
column $\ep_1$, acting on space $V=\alpha,i$ (site $i$ tensor copy $\alpha$) or
$V={\rm aux}$ (auxiliary space). For the quantum state $\psi_{\{a_{\vec{v}}\}}$, we write
\[
\psi_{\{a_{\vec{v}}\}}^{\otimes n} = \sum_{\{\vec{v}_\alpha:\alpha=1,\ldots,n\}}
\lt(\prod_{\alpha=1}^n |a_{\vec{v}_\alpha}|^2\rt)
        \bigotimes_{\alpha=1}^n \psi_{\vec{v}_\alpha}.
\]
From the trace expression (\ref{trace}) we find \beq
     \bigotimes_{\alpha} \psi_{\vec{v}_\alpha} \lt({\cal T}_A\rt) =
        \prod_{i\in A}\text{Tr}_{\text{aux}} \prod_{\alpha} \sum_{\ep_1,\ep_2} E_{\text{aux}}^{\ep_1\ep_2}
        \psi_{\vec{v}_\alpha;i}\lt(E_{i}^{\ep_2\ep_1}\rt).
\eeq Clearly, $\psi_{\vec{v}_\alpha;i}\lt(E_{i}^{\ep_2\ep_1}\rt)$
is independent of $i$. Writing $|\psi_{\vec{v}}\ket =
s_{\vec{v},1}|\uparrow\ket + s_{\vec{v},2}|\downarrow\ket$, we
find $\psi_{\vec{v}_\alpha,i}\lt(E_{\alpha,i}^{\ep_2\ep_1}\rt) =
s_{\vec{v}_\alpha,\ep_2}^* s_{\vec{v}_\alpha,\ep_1}$, and tracing over the auxiliary space we obtain
\[
    \text{Tr}_{\text{aux}} \prod_{\alpha} \sum_{\ep_1,\ep_2} E_{\text{aux}}^{\ep_1\ep_2}
        \psi_{\vec{v}_\alpha;i}\lt(E_{i}^{\ep_2\ep_1}\rt) =  \prod_{\alpha} \bra\psi_{\vec{v}_\alpha}|\psi_{\vec{v}_{\alpha+1}}\ket.
\]
Hence, we find \beq\label{rgsa}
    S_n= \frc1{1-n} \log\lt(\sum_{\{\vec{v}_\alpha\}}
        \lt[\prod_{\alpha} |a_{\vec{v}_\alpha}|^2 \rt]
        \lt[\prod_{\alpha} \bra\psi_{\vec{v}_\alpha}|\psi_{\vec{v}_{\alpha+1}}\ket\rt]^{m}\rt)
\eeq
with $\vec{v}_{n+1} := \vec{v}_1$. This saturates at large
$m$ to \beq\label{saturation}
    \lim_{m\to\infty} S_n =
    \frc1{1-n}\log\lt(\sum_{\vec{v}} |a_{\vec{v}}|^{2n}\rt).
\eeq That is, as expected,
 for any ground state given by a finite linear combination of
 basic zero entropy states, the entanglement entropy reaches a finite maximum as
 the number $m$ of site of $A$ tends to infinity. This corresponds to
 the case $d=0$ in (\ref{main}).

More interesting behaviours are obtained from ``infinite linear combinations'' of basic states, generalising (\ref{pa}). Given a smooth, self-avoiding path $\vec\gamma:[0,1]\to
S^2$ and a smooth function $f:S^2\to\R^+$ with
$\int_0^1 |d\vec{\gamma}(t)|\,f(\vec{\gamma}(t)) = 1$, the
following integral can be defined and is a quantum ground state:
$\psi^{(1)}:= \int_0^1 |d\vec{\gamma}(t)|  \,f(\vec{\gamma}(t))\, \psi_{\vec\gamma(t)}$.
Similarly, let $\vec\mu:[0,1]\times [0,1] \to
S^2$ be a two-dimensional smooth curve such that $\int_0^{1} \int_0^{1} |d^2\vec{\mu}(\lambda,\phi)|
    f(\vec{\mu}(\lambda,\phi))=1$ (where
$|d^2\vec{\mu}(\lambda,\phi)|$ is the surface element on the unit sphere). Then, the following is a ground state:
$
 	\psi^{(2)}:= \int_0^{1} \int_0^{1} |d^2\vec{\mu}(\lambda,\phi)|\, f(\vec{\mu}(\lambda,\phi))\,
    \psi_{\vec\mu(\lambda,\phi)}
$.  Generalising, we may consider the set of
non-zero coefficients to be a
fractal set $\mathcal{W} \subset S^2$ with fractal
dimension $d$. With $d{\mathcal{H}}(\vec{v})$ the corresponding Hausdorff
integration measure,
we may write
\[
\psi^{(d)}:=\int_{\mathcal{W}}d{\mathcal{H}}(\vec{v})
\,f(\vec{v})\, \psi_{\vec{v}} \,\, \,\text{with}\,\,
\int_{\mathcal{W}}d{\mathcal{H}}(\vec{v})\, f(\vec{v})= 1.
\]
The Hausdorff measure is expected to occur naturally in taking the
large-volume limit,
if the set of vectors $\vec{v}$ such that $a_{\vec{v}}\neq0$ becomes
a fractal set.

Computing the R\'enyi entropy of $\psi^{(d)}$ yields a simple generalisation of (\ref{rgsa}) where
the sums over the vectors $\vec{v}_\alpha$ are replaced by
integrations and the coefficients $|a_{\vec{v}_\alpha}|^2$ by the
functions $f(\vec{v}_\alpha)$. The logarithmic factor in (\ref{rgsa}) becomes
\beq\label{rgsf}
    \log\lt(\int_{\mathcal{W}} \prod_{\alpha}
     d{\mathcal{H}}(\vec{v}_\alpha)\,f(\vec{v}_\alpha)
        \lt(\prod_{\alpha} \bra\psi_{\vec{v}_\alpha}|\psi_{\vec{v}_{\alpha+1}}\ket\rt)^{m}\rt).
\eeq The large $m$ asymptotics of these expressions will however
be radically different from (\ref{saturation}): there will
be no saturation, and we will recover the behaviour
highlighted in (\ref{main}). This can be shown using a saddle-point analysis,
as was done in \cite{permutation} in a particular case; here it is generalised
to integrals over fractal domains.

For the explicit calculations, we use $
    |\psi_{\vec{v}}\ket = \frc1{\sqrt{2}} \mato{c} \sqrt{1+z} \\ \sqrt{1-z} \, e^{i\theta}\matf
$, where $\vec{v}=:(x,y,z)$ is a unit vector, and $\theta$ is defined by $x+iy
= \sqrt{1-z^2}\, e^{i\theta}$.
From this we see that $|\bra\psi_{\vec{v}}|\psi_{\vec{w}}\ket|\leq1$, with equality
if and only if $\vec{v}=\vec{w}$. The saddle-point analysis from (\ref{rgsf}) is done by expanding
the overlaps $\bra\psi_{\vec{v}_\alpha}|\psi_{\vec{v}_{\alpha+1}}\ket$ around $\vec{v}_\alpha = \vec{v}_{\alpha+1}$,
and re-writing the product $\prod_{\alpha}$ of these overlaps as an exponential. We get
\beq
    \prod_{\alpha=1}^n \bra\psi_{\vec{v}_\alpha}|
    \psi_{\vec{v}_{\alpha+1}}\ket
    = \exp\lt[-\frac{1}{8}\sum_{\alpha=1}^n |\vec{v}_{\alpha+1}
    -\vec{v}_\alpha|^2 + \ldots\rt],\label{eq1}
\eeq
where the ellipsis stand for terms that are order-2 and antisymmetric,
and higher order terms. The order-2 antisymmetric terms vanish
when the integrations in (\ref{rgsf}) are performed.

In the case of $\psi^{(1)}$ for instance, we may
use the assumptions relating to $f$ and $\gamma$: both are smooth,
and the curve $\gamma$ is self-avoiding. Hence, with $\vec{v}_\alpha = \vec{\gamma}(t_\alpha)$, the maximum
occurs when $t_\alpha = t_{\alpha+1}$ for all
$\alpha=1,\ldots,n$. In this case we find \beq
    \prod_{\alpha=1}^n \bra\psi_{\vec{\gamma}(t_\alpha)}|
    \psi_{\vec{\gamma}(t_{\alpha+1})}\ket
    = \exp\lt[-\frac{|\dot{\vec{\gamma}}|^2}{8}\sum_{\alpha=1}^n (t_{\alpha+1}
    -t_\alpha)^2 + \ldots\rt],\label{eq2}
\eeq where $|\dot{\vec{\gamma}}|^2$ is evaluated at $t=t_1$.
The saddle-point analysis is then performed as follows.
We need to raise the quantity above to the power $m$ and
substitute into the integral (\ref{rgsf}). We can replace $f(\vec\gamma(t_\alpha))$ by $f(\vec\gamma(t_1))$ for
all $\alpha$, since $f$ is smooth. Changing variables to $\hat{t}_i=\sqrt{m}({t_{i}-t_{1}})$, $i=2,\ldots,n$
guarantees that larger positive powers of $\h{t}_\alpha$ give
lower-order contributions at large $m$. We obtain
\begin{eqnarray}
   S_n&=& \frc1{1-n}
    \log\lt(\frac{1}{m^{\frac{n-1}{2}}}\int_{0}^{1} dt_1|
    \dot{\vec{\gamma}}(t_1)|^n
    f(\vec{\gamma}(t_1))^n \rt.\label{int}\\
    && \lt. \int_{-\infty}^{\infty}
    d^{n-1}\hat{t} \,
    e^{-\frac{|\vec{\gamma}|^2}{8}\left[\sum\limits_{\alpha=2}^{n-1}
    (\hat{t}_{\alpha+1} - \hat{t}_{\alpha})^2+\h{t}_2^2 + \h{t}_n^2
     \right]+ O(\hat{t}^3/\sqrt{m})}\rt),\nonumber
    \end{eqnarray}
where the integrals over $\h{t}_2,\ldots,\h{t}_n$ have been extended to
$(-\infty,\infty)$ (the resulting correction terms are exponentially small).
These integrals are of standard gaussian type
and can be carried out explicitly.

A very similar computation can be carried out for the state
$\psi^{(2)}$ instead of $\psi^{(1)}$. The final result can be expressed
in both cases $d=1$ and $d=2$ as
\beq\label{logm}
    S_n \sim \frc{d}2 \log \lt(\frc{m}{8\pi}\rt)  +\frac{1}{1-n}
    \log\lt(n^{-\frc{d}2}\int |d^d\vec{v}|\,
    \,f(\vec{v})^n\rt),
    \eeq
where higher order corrections would be $O(m^{-1/2})$. This is in agreement with (\ref{main}).

Note that the constant term in (\ref{logm}) is $-\log(f_{\rm max})$ as $n\to\infty$, where $f_{\rm max}$ is the maximum of $f$ on its support. At $n=1$, we have rather $d/2-\int |d^d\vec{v}|\, f(\vec v) \log f(\vec{v})$.

The calculation for fractal sets follows similar lines. A crucial feature of
the Hausdorff measure is its scaling property. On the plane, the Hausdorff measure ${\cal H}'$ satisfies
$
     s^{d} \, {\mathcal{H}}'(\mathcal{W}') ={\mathcal{H}'}(s
   \mathcal{W}'+\vec{u})
$
for any $\mathcal{W}'\subset\R^2$. For the measure $\mathcal{H}$ on $S^2$, this scaling covariance is replaced by
an asymptotic behaviour that gives rise to the measure $\mathcal{H}'$ on the tangent plane:
\beq\label{asympt}
	\lim_{m\to\infty} m^{d/2}\,d{\cal H}(\h{\vec{v}}_i/\sqrt{m} + \vec{v}_1) = d{\cal H}'(\h{\vec{v}}_i).
\eeq
Putting (\ref{eq1}) inside (\ref{rgsf}), changing variables to
$\h{\vec{v}}_i = \sqrt{m}(\vec{v}_i-\vec{v}_1)$, $i=2,\dots,n$, and using (\ref{asympt}), as $m\to\infty$,
\begin{eqnarray}
   S_n&\sim& \frc1{1-n}
    \log\lt(\frac{1}{m^{\frac{d(n-1)}{2}}}\int_{{\cal W}} d{\cal H}(\vec{v}_1)
    f(\vec{v}_1)
    \int_{{\cal W}_m'} \prod_{i=2}^n 
     \rt.\label{intd}\n
    && \lt. \hspace{-1cm}  d{\cal H}'(\h{\vec{v}}_i)
    f\lt(\frc{\h{\vec{v}}_i}{\sqrt{m}}
    +\vec{v}_1\rt) \,
    e^{-\frac{1}{8}\left[\sum\limits_{\alpha=2}^{n-1}
    |\hat{\vec{v}}_{\alpha+1}-\hat{\vec{v}}_{\alpha}|^2+|\hat{\vec{v}}_2|^2 + |\hat{\vec{v}}_n|^2
     \right]}\rt)\nonumber
    \end{eqnarray}
where ${\cal W}_m' = \sqrt{m}({\cal W}-\vec{v}_1)$ (projected to the tangent plane at $\vec{v}_1$). Although the integral might not exist in the large-$m$ limit, it is bounded, thanks to the exponentially decaying factor. This boundedness immediately gives rise to the leading
asymptotics (\ref{main}).
It would be desirable to investigate more precisely the nature of the
constant corrections to this general leading behaviour; we hope to return to this
in a future work. {\bf Acknowledgment:} We would like to thank J.L. Cardy for useful comments.


\end{document}